\documentstyle[epsf,aasms]{article}
\tighten
\begin{document}

\title{On the oscillation spectra of ultracompact stars:
An extensive survey of gravitational-wave modes}

\lefthead{Andersson, Kojima and Kokkotas}
\righthead{Survey of gravitational-wave modes}

\author{Nils Andersson}
\affil{Department of Physics and Astronomy,
University of Wales College of Cardiff, CF2 3YB, United Kingdom\\
Department of Physics, Washington University, St Louis MO 63130, USA}

\author{Yasufumi Kojima}
\affil{Department of Physics, Hiroshima University,
Higashi-Hiroshima 739, Japan }

\author{Kostas D. Kokkotas}
\affil{Department of Physics, Aristotle University of Thessaloniki,
Thessaloniki 54006, Greece}

\abstract
An extensive survey of gravitational-wave modes for uniform density stars
is presented. The study covers stars
ranging in compactness from $R/M=100$ to the limit of stability in
general relativity: $R/M = 9/4$. We establish that polar and axial
gravitational-wave modes exist for all these stellar models.
Moreover,  there are two distinct families of both axial and polar modes.
We discuss the physics of these modes and argue that one family
is primarily
associated with the interior of the star, while the second family is
mainly associated with the stellar surface.
We also show that the problem of axial perturbations has all the
essential features of the polar problem as far as  gravitational waves
are concerned. This means that the axial problem is much more important
than has previously been assumed. We also find some surprising features, such
as avoided crossings between the polar gravitational-wave modes and
the Kelvin f-mode as the star becomes very compact.
This seems to suggest that the f-mode should be
 considered on equal footing with the polar w-modes for ultracompact stars.
All modes may have the main character of trapped modes inside
the curvature potential barrier for $R/M < 3$.
\endabstract


\section{Introduction}

Neutron stars have remarkably rich oscillation spectra (McDermott, Van
Horn and Hansen 1988). Essentially every characteristic of the star --
such as its density, temperature, rotation, magnetic field et cetera --
can be, more or less directly, associated with a unique set of
oscillation modes. This makes a study of the full problem of pulsating
neutron stars hopelessly difficult. The only reasonable  approach is to
strip off all the features that are not expected  to interfere with the
principal physical mechanism behind a certain pulsation mode.  Once the
``simple'' situation is explored one can add, one by one, other
features of the star back on and (possibly) verify that the
approximation made was a valid one.   This approach has been successful
in the past, and our understanding of  pulsations that are mainly
associated with the stellar fluid has been much improved.

The oscillation modes of a star can be divided into two general
classes; polar (spheroidal, or even parity)
and axial (toroidal, or odd parity). The polar ones
correspond to spheroidal deformations of the star, whereas the axial
ones are associated with differential rotation. As far as the stellar
fluid is concerned, polar pulsation modes exist for all conceivable
stellar models, whereas the existence of axial modes rely upon nonzero
rotation, magnetic field or shear modulus. To set the scene for this
paper, let us discuss a hierarchy of increasingly  complex stellar
models and the pulsation modes associated with each of them;

\begin{enumerate}
\item The simplest possible model for a neutron star is a non-rotating ball
 of fluid at zero temperature. If the fluid has constant density this
model can only support a single pulsation mode for each multipole
$\ell$. (The oscillations are typically described in terms of the
standard spherical harmonics $Y_{\ell m}$.) This  mode, which in Newtonian
theory has
oscillation frequency
\begin{equation}
 \omega (R^3/M)^{1/2}= \sqrt{ 2l (l-1)/(2l+1) }
 \approx 0.894 \quad {\rm for}\ \ell=2 \ ,
\label{fmode}\end{equation}
where $R$ and $M$ are the radius and mass of the star in units where
$c=G=1$, was first studied by Kelvin (Cox 1980) and is usually referred
to as the f-mode
 (fundamental mode) (Cowling 1941).  It is distinguishable by the fact
 that there are no nodes in the corresponding eigenfunctions inside the
 star. In a way, the f-mode is due to the interface between the star
 and its surroundings. The eigenfunctions of such modes would typically have
 maxima at the interface and fall off away from it (McDermott, van Horn
 and Hansen 1988). This is exactly the character of the f-mode: It
 reaches maximum amplitude at the surface of the star (see for example
 Figure 7--8 in Kokkotas and Schutz 1992).

\item A somewhat more realistic star consists of a perfect fluid. Then one
must specify the equation of state, and most studies to date have been
for simple polytropes. For this stellar model a second set of
modes --  the p-modes (Cowling 1941),  the restoring force of which is
pressure -- exists. The  oscillation frequencies of the p-modes depend
directly on the travel time for an acoustic wave across the star. This
would typically lead to an oscillation period shorter than a few
milliseconds.

\item  When the temperature of the star is non-zero a further set of modes
 come into play. These modes are mainly restored by gravity and are
 consequently referred to as g-modes (Cowling 1941). For a star in convection,
 {\em i.e.,} when the entropy is constant, the g-modes are all degenerate at
 zero frequency. In general, however, their oscillation
 periods depend directly on the central temperature of the star and are
 typically longer than a few hundred milliseconds.

\item The three families of ``fluid'' modes discussed so far, the f-, p- and
 g-modes, all belong to the class of polar modes. For these models there are no
 ``fluid'' axial modes. But then, the
 stellar models discussed are all somewhat unrealistic. In reality, it is
 expected that the crust of the star will  crystallize, and a typical
 neutron star would have a kilometer-thick crust. When
 the shear modulus in the crust is nonzero axial modes exist (Schumaker
 and Thorne 1983). There will also be families of modes directly
 associated with the various interfaces in the star (McDermott, van
 Horn and Hansen 1988).

\end{enumerate}

Although a real neutron star will most likely be rapidly rotating, have
a strong magnetic field, contain a region of superfluidity et cetera we
will not discuss these complications here. What we want to emphasize
with the above examples  is that {\em each feature of the star is
associated with at least one unique set of pulsation modes}. The more
complicated the stellar model, the richer is its oscillation spectrum.

The picture painted so far is nevertheless not satisfactory since we
have not included mechanisms for dissipation. Pulsations of a real
neutron star will be damped. For example, displacements close to the
surface of the star will affect the external magnetic field, and
``shaking''  magnetic field lines will generate electromagnetic waves.
These can carry  pulsation energy away from the star. For pulsars this
might result in observable perturbations of the pulsar beam (McDermott,
van Horn and Hansen 1988). A second dissipation mechanism is due to
general relativity. Gravitational waves will be generated when the
stellar fluid sloshes about. Emission of gravitational waves is the
dominant dissipation mechanism for many stellar modes.

The (relativistic) Cowling approximation (Cowling 1941; McDermott, Van
Horn and Scholl 1983) has been an invaluable tool for achieving the
present understanding of neutron star pulsations. In this approximation
all perturbations of the gravitational field are neglected, and the
pulsation equations simplify considerably. The damping due to
gravitational waves can be estimated by means of the celebrated
quadrupole formula (Balbinski and Schutz 1982; Balbinski {\em et al.} 1985).
This leads to a typical
damping time of a few seconds for the f- and the p-modes. Compared
to this, the gravitational-wave damping of the g-modes is incredibly
slow. Very little mass is involved in the g-mode pulsations -- the
modes are localized close to the surface (or the centre) of the star,
and  the damping time is estimated to be of the order of the lifetime
of the star. For the g-modes the damping due to ``shaking'' of the
magnetic field is more important.

Calculations for perturbed stars in general relativity were pioneered
by Thorne and Campolattaro in 1967. Since then many authors have
considered the problem, but there are not many numerical studies of
mode-frequencies. The most extensive one, due to Lindblom and Detweiler
(1983), considers many different equations of state but only the
quadrupole f-mode for each star. Until a few years ago the consensus
was that general relativity would not considerably change the results
of Newtonian theory. Each mode-frequency would only adopt a small
imaginary part (making it complex) to account for damping due to
gravitational waves.

It is now becoming clear, however, that general relativity is playing a
much more exciting role. According to Einstein's theory the
gravitational field should be considered a dynamical entity, and
neutron stars are very compact objects for which relativity should be
of some importance. Hence, it seems likely that a neutron star will
have pulsation modes mainly associated with the gravitational field
itself: A kind of ``spacetime'' modes that are, in some sense, coupled
to the ``fluid'' modes of Newtonian theory. These modes would be
reminiscent of the quasinormal modes of black holes (Chandrasekhar 1983).

 That such modes should exist was first argued by Kokkotas and Schutz
 (1986). They suggested a simple model problem of a finite
 string -- representing the stellar fluid -- coupled by means of a
 spring to a semi-infinite string -- the gravitational field.  This system
 supports two sets of pulsation modes; one that
 is slowly damped and analogous to the ``fluid'' modes of a star, and a
 second one that is rapidly damped and {\em has no analogue in
 Newtonian theory}. This new set of modes has  been termed w-modes
 because of the close association to gravitational waves. That such
 modes exist also for polytropes has since been demonstrated by several
 studies (Kojima 1988; Kokkotas and Schutz 1992; Leins, Nollert and
 Soffel 1993; Andersson, Kokkotas and Schutz 1995a).

The problem of axial perturbations of stars has not attained much
interest in the past. The main reason for this is that axial
perturbations do not couple to oscillations of the stellar fluid. As
was shown by Thorne and Campolattaro in their seminal paper of 1967,
axial perturbations are governed by a single homogeneous wave equation
for one of the perturbed metric functions. For three of  the simple
stellar models discussed above there are no axial pulsation modes
associated with the fluid. In 1983 Schumaker and Thorne developed a
general relativistic description of axial modes for a neutron star with
a crust. This is one of very few discussions of axial modes in the
literature.

An exciting idea of Chandrasekhar and Ferrari (1991b) recently
brought the axial modes more into focus. If the star is made very
compact -- roughly when $R/M<3$ -- the problem for axial modes becomes
similar to that for quasi-bound scattering resonances in quantum
mechanics. That is, one gets an effective potential with a well inside
the barrier that is familiar from black-hole problems (Chandrasekhar
1983). Clearly, axial modes associated with this potential well should
exist and Chandrasekhar and Ferrari found a few such
modes. Since then, results of Kokkotas (1994) have indicated that there
may, in fact, exist an infinite number of axial modes for these compact
stars. Most of these modes are rapidly damped, which explains why they could
not be identified with the numerical technique employed by
Chandrasekhar and Ferrari.

It is interesting to note here that, the idea of trapped waves for very
compact stars is not at all new. That the black-hole potential barrier
may affect the f-mode for sufficiently compact stars was first pointed
out by Detweiler (1975). Moreover, Vishveshwara and his co-workers
discussed ultracompact stars in a series of interesting papers. The
first of these (Kembhavi and Vishveshwara 1980), concerns neutrinos
trapped by neutron stars. This problem is remarkably similar to that
for axial modes and compact stars; there is a well inside a potential
barrier. Two following papers (Iyer and Vishveshwara 1985; Iyer,
Vishveshwara and Dhurandhar 1985), investigate whether ultracompact
stars may exist for ``realistic'' neutron star equations of state.
 It is clear that one can only have $R/M<3$ for extremely stiff
equations of state. Nevertheless, using the so-called core-envelope
model, the authors show that stable ultracompact objects with causal
cores may exist. On the other hand, no such objects were found using
the available high-density equations of state.

The present paper mainly concerns gravitational-wave modes for uniform density
stars. We present the most extensive survey of such modes carried out
so far. Admittedly, we have chosen an unrealistic stellar model, but we
have good reasons for doing so.  The main advantage is that we can
study a sequence of models with continuously varying surface redshift.
That is, it is straightforward to do calculations covering all degrees
of compactness,  including white dwarfs: $R/M \sim 10^4$, neutron
stars:  $R/M \sim 5$ and continuing towards the ultimate limit of
relativistic stability at $R/M = 9/4$. By following individual modes as
the stellar models change in this way one may hope to understand better
the physical origin of the modes.

Another motivation behind the present study is completeness. We have
previously considered the problem for modes that are slowly damped
(Kojima, Andersson and Kokkotas 1995), but only for stars close to the
ultimate limit posed by general relativity. We wanted to see whether
the axial and the polar modes behaved in a similar way as this limit
was approached. That was expected since both sets would then behave as
``trapped'' modes inside very similar potential barriers. That the two
sets do ``approach'' each other as $R/M\to 9/4$ was, indeed, one of the
conclusions of our previous work.
But that study was limited in several ways by the technique we used to
identify polar modes; the so-called resonance method. Inherent in that
technique is that it will only work as long as the imaginary part of a
mode-frequency is much smaller than the real part (Chandrasekhar,
Ferrari and Winston 1991), and hence we could not hope to study rapidly
damped w-modes with it. Furthermore, our results indicated that the
w-modes would cross the f-mode for very compact models. Again due to
the nature of the resonance method, we could not resolve these possible
mode-crossings.

Despite the restrictions, the results of our previous work were very
interesting. It seemed clear that the origin of the axial and the polar
modes \underline{must} be the same for these very compact models.  If
so, it  seems reasonable that both sets are due to the same
physical mechanism in general; that they are ``spacetime'' modes that
{\em do not depend on the fluid at all for their existence}. This does
not mean that the fluid would not affect the polar modes. It certainly
will, but {\em the modes exist even if we freeze the fluid motion}
(Andersson, Kokkotas and Schutz 1995b). If this idea is correct, the
existence of rapidly damped axial gravitational-wave modes for
\underline{all} stellar models is unavoidable.
In this context it is relevant to point out that the established  idea
that axial modes will not exist for less compact stars relies heavily
on Newtonian theory where the fluid plays the main dynamic role. It is
quite easy to argue that axial perturbations should be as relevant as
the polar ones as far as the gravitational field is concerned. Evidence
for this is certainly provided by the black-hole problem, where the
axial and polar spectra are identical (Chandrasekhar 1983). The present
study aims to establish the existence of axial gravitational-wave modes
for all stellar models.

\section{On the problem}

The equations governing perturbed stars in general relativity have been
considered in great detail by many authors. Thus, we do not find it
necessary to go into much detail here. We will simply refer the
interested reader to the original sources for the exact form of most
equations. Nevertheless, an understanding of the general form of
some of these equations will be useful for a discussion of the present
results and  we introduce the necessary material in this section.  We
will also comment on  our present numerical work and assess its
reliability.

\subsection{Perturbed stars in general relativity}

We use the unperturbed metric
\begin{equation}
ds^2 = -e^\nu dt^2 + e^\lambda dr^2
+ r^2 \left( d\theta^2 +\sin^2 \theta d\phi^2 \right) \ ,
\label{unpert}\end{equation}
where $\nu$ and $\lambda$ are functions of $r$ only. In fact,
\begin{equation}
e^{-\lambda} = 1 - {2m(r)\over r} \ ,
\end{equation}
where $m(r)$ represents the mass inside radius $r$. For a fluid ball
 of uniform density it is easy to show that
\begin{equation}
m(r) = M \left( { r \over R} \right)^3 \ ,
\end{equation}
inside the star. The total mass of the star is $M = m(R)$. The
metric variable $\nu$ is determined by the equations of hydrostatic
equilibrium, and for uniform density we get (Schutz 1985)
\begin{equation}
e^\nu =  {1\over 4} \left[ 3 \left( 1 - {2M \over R} \right)^{1/2} -
\left( 1 - {2m(r) \over r} \right)^{1/2} \right]^2 \ .
\end{equation}
 In the vacuum outside the star this should be replaced by
\begin{equation}
e^\nu =  \left( 1 - {2M \over r} \right) \ ,
\end{equation}
and the metric becomes the standard Schwarzschild metric.

We now wish to consider linear perturbations of (\ref{unpert}). That is,
we introduce a perturbed metric
\begin{equation}
g_{\mu \nu} = g_{\mu \nu}^{\rm background} + h_{\mu \nu}^{\rm polar} +
h_{\mu \nu}^{\rm axial} \ ,
\end{equation}
where $h_{\mu \nu}$ are (in some sense) small. As we have already
stated, the polar perturbations correspond to spheroidal deformations
of the star, whereas the axial ones correspond to differential
rotation. The equations governing these two classes will decouple. We
also need variables describing the perturbations of the fluid. We
could, for example, use displacement components $\xi^r$ and
$\xi^\theta$ together with the Eulerian variation in the density
$\delta \rho$. We essentially have thirteen undetermined functions. The
equations that govern the evolution of the perturbed metric and the
stellar fluid are
(i) ten perturbed Einstein equations, and (ii) three equations of
motion for the fluid. Clearly, this is a well posed problem. However,
because of the Bianchi identities only ten of our variables will be
independent. For example, if we decide to use the perturbed Einstein
equations to determine all ten metric perturbations the three equations
of motion for the fluid will  automatically be satified. Hence, several
different sets of perturbation equations can be used, depending on ones
favourite variables. The possibilities also involve various gauge choices.

\subsection{The perturbation equations}

Although it is by no means clear what the ``best'' choice of variables
may be,  it is straightforward to prescribe the expected form of the
final perturbation equations. In the case of axial perturbations the
expectations are, in fact, easily realized since these perturbations do
not couple to oscillations of the fluid. Three of the perturbed metric
components are axial, and as was shown by Thorne and Campolattaro
(1967) one of these is governed by an equation of form
\begin{equation}
-{\partial^2 X \over \partial t^2}
+ {\partial^2 X \over \partial r_\ast^2} - V X = 0 \ ,
\label{waveq}\end{equation}
where the tortoise coordinate $r_\ast$ is defined by
\begin{equation}
{\partial \over \partial r_\ast} = e^{(\nu-\lambda)/2}
{\partial \over \partial r} \ .
\end{equation}
The exact form of the function $X=X(h_{r \theta})$ and
the potential $V$ are not of great concern to us here. They are given
explicity by, for example, Chandrasekhar and Ferrari (1991b) who also
graphed the potential for  a few very compact uniform density stars.
Some general properties of the potential are useful to keep in mind,
however;  $V$ is positive for all $r$, and in the exterior vacuum it
reduces to the Regge-Wheeler potential (Chandrasekhar 1983). This
potential reaches a maximum at $R/M \approx 3$ and vanishes as $1/r^2$
as $r\to \infty$.  Inside the star, the potential diverges as
$\ell(\ell+1)/r^2$ as $r \to 0$.

In the exterior vacuum, the polar perturbation problem also simplifies
to a single equation of form (\ref{waveq}). This is the so-called
Zerilli equation (Chandrasekhar 1983). The polar effective potential is
similar, albeit not identical, to that for the axial case. The description
of the polar perturbation problem inside the
star is not as
straightforward. Several different sets of equations have been derived
and used. It has long been accepted that the problem is one of fourth
order. For example, Lindblom and Detweiler (1983) used Regge-Wheeler
gauge and derived four coupled first-order equations for $[h_{tt},
h_{rr}, \xi^r, \xi^\theta]$. This system was used in their numerical
study of f-modes for many different equations of state. They later
realized (Detweiler and Lindblom 1985) that this system of equations
could become singular inside the star and that it was preferable to use
$h_{tr}$ instead of $h_{tt}$.
Recently, Chandrasekhar and Ferrari (1991a) approached the problem in
diagonal gauge and derived a fifth-order system of equations.
Interestingly, their new set of equations does not explicitly contain the
fluid variables and hence the problem of gravitational waves scattered
off a star can be considered in much the same way as that for black
holes. Price and Ipser (1991) have shown that
the Chandrasekhar-Ferrari equations can also be reduced to fourth
order. The fifth equation corresponds to a solution that is ruled out
by physical requirements and is thus superfluous.
A fourth-order system of equations without explicit dependence on the
fluid variables in Regge-Wheeler gauge was first derived Ipser and Price
(1991). It has been extended to the case of slow rotation
by Kojima (1992,1993).
 In our opinion, these latter equations are the most appealing
ones derived so far. They consist of two coupled wave equations, corresponding
to the two dynamical degrees of freedom; the fluid oscillations and the
gravitational waves. The equations for a barotropic case are given by
\begin{equation}
- {1\over C^2} {\partial^2 Y \over \partial t^2} +
{\partial^2 Y \over \partial r_\ast^2} + F \left(
{\partial Y \over \partial r_\ast} ,
{\partial Z \over \partial r_\ast} , Y, Z \right) = 0 \ ,
\end{equation}
\begin{equation}
- {\partial^2 Z \over \partial t^2} +
{\partial^2 Z \over \partial r_\ast^2} + G \left(
{\partial Y \over \partial r_\ast} ,
{\partial Z \over \partial r_\ast} , Y, Z \right) = 0 \ ,
\end{equation}
where $C^2 = \delta p / \delta \rho$ and the functions $Y$ and $Z$ are linear
combinations of the metric perturbations $h_{tt}$ and $h_{\theta \theta}$.
These equations clearly represent the hydrodynamical and gravitational
wave propagation. In this paper, we adopt the approximation $\delta \rho = 0$,
or equivalently $C^2 \to \infty$. Then the basic equations become
one constraint equation and one wave equation. There are no singularities
in the equations introduced in this approximation since they include only
$1/C^2$ and not $C^2$.

\subsection{Comments on the present work}

The quasinormal modes of a stellar system are solutions to the
perturbation equations that satisfy the physical condition of
regularity at the centre of the star, appropriate matching conditions
at the surface and also correspond to purely outgoing gravitational
waves at infinity. Let us assume that standard Fourier-decomposition
leads to solutions to (\ref{waveq}) that depend on time as
$\exp(i\omega t)$. In the following we will refer to
the corresponding time-independent equation as (\ref{waveq}).
It follows, since both the
axial and the polar potential vanishes as $r\to \infty$, that a
quasinormal mode characterised by a frequency $\omega_n$ is a  solution
to (\ref{waveq}) that behaves as
\begin{equation}
X \sim e^{-i\omega_n r_\ast} \quad {\rm as}\ r_\ast \to \infty \ .
\label{outgoing}\end{equation}
Identifying such a solution is a far from a trivial problem.
We expect the modes of a star to be damped as gravitational waves are
emitted. This means that $\omega_n$ must be complex with a positive
imaginary part. An observer sitting at a constant value of $r_\ast$
will then see a gravitational wave that is exponentially damped with time.
The boundary condition (\ref{outgoing}) thus requires that we identify a
solution that diverges as $r_\ast \to \infty$. This is clearly not easy
to do since we must ensure that no  (exponentially small) ingoing waves
remains at infinity.

In our previous study, of slowly damped modes for ultracompact stars
with uniform density, we used the so-called resonance method to
identify polar modes. In this method one constructs the ratio of the
asymptotic amplitudes for out- and ingoing waves for \underline{real}
values of $\omega$. The quasinormal modes with small imaginary part can
then be identified as singularities [see Figure 1 in Kojima, Andersson
and Kokkotas (1995)]. This method clearly fails for rapidly damped
modes and is therefore not useful for a survey of the present kind. In
a study of rapidly damped modes the frequency must be considered
complex.

The divergence-problem is well-known from black-hole studies and
several ways to overcome it have been developed. Some of these have
also been extended to the problem of rapidly damped stellar modes.
Kokkotas and Schutz (1992) used a WKB approximation at the surface of
the star, whereas Leins, Nollert and Soffel (1993) employed Leaver's
continued fraction method (Leaver 1985) as well as the Wronskian scheme
developed by Nollert and Schmidt (1992). Here we employ a
complex-coordinate approach that was developed for black holes by
Andersson (1992) and recently applied to the stellar problem by
Andersson, Kokkotas and Schutz (1995a). The general idea behind this
scheme is very simple. By allowing the coordinate $r_\ast$ to assume
complex values we can suppress the divergence at $ \vert r_\ast \vert
\to \infty$. Specifically,
for each complex frequency $\omega_n$ one can always find a contour in
the complex $r_\ast$-plane such that the out- and ingoing waves, {\em
i.e.}, the two asymptotic solutions $\exp(-i\omega_n r_\ast)$
and $\exp(i\omega_n
r_\ast)$ to (\ref{waveq}), are of the same order of magnitude.
Along such a contour it is straightforward to ensure that the
outgoing-wave boundary condition is satisfied. In fact, since the
potentials vanishes as $\vert r_\ast \vert \to \infty$ it is clear that
such contours will asymptotically be straight lines with a slope equal
to $-{\rm Im}\ \omega_n / {\rm Re}\ \omega_n$ . In order to use this
method one must, of course, prove that a boundary condition introduced
for complex $r_\ast$ actually corresponds to the desired behaviour on
the real $r_\ast$-axis -- where the physical condition is imposed. The
simplest argument for this appeals to analyticity. This hand-waving
argument has been supported by a semiclassical demonstration by
Ara\'ujo, Nicholson and Schutz (1993).

The method developed by Andersson, Kokkotas and Schutz
(1995a) must be used with some care for ultracompact stars, however. This
is an important point and we will try to explain it without going into
too much detail here.  Solutions to (\ref{waveq}) are constructed as linear
combinations of two numerically determined functions which
asymptotically become equal to the two WKB solutions, and
thus represent  out- and  ingoing waves at infinity.
These funcions can be split into an amplitude and a phase, which are
related because the Wronskian of any two linearly independent
solutions to (\ref{waveq})
\underline{must} be constant. Hence, it is sufficient to consider the
phase-function. In the Andersson-Kokkotas-Schutz method this
phase-function is numerically determined along a straight line in the
complex $r$-plane. This contour ends at the surface of the star, and is
chosen such that the out- and ingoing wave solutions are of the same
order of magnitude far away from the star, {\em i.e.,} with a slope
$-{\rm Im}\ \omega_n / {\rm Re}\ \omega_n$.

With this method the validity of the geometrical optics argument -- that
waves are not backscattered by the curvature in the exterior of the
star -- that is the essence of the WKB approximation that Kokkotas and
Schutz (1992) used in their study of w-modes, can be tested. It was
shown (Andersson, Kokkotas and Schutz 1995a), that the geometrical
optics argument is perfectly valid for most w-modes of polytropes.
However, if wave-reflection by the exterior curvature becomes
considerable, for example when the star becomes very compact and the
black-hole potential barrier in unveiled (as $R/M \approx 3$),
geometrical optics will clearly no longer be reliable.

It can be shown that the phase-function that is the key quantity in the
Andersson-Kokkotas-Schutz method will also be affected as waves are
backscattered by the curvature. It is quite obvious that it must,
since all information
about the exact solution to (\ref{waveq}) will be encoded in the
phase-function. The function is generally smooth and well behaved --
which makes it ideally suited for numerical studies -- as long as the
wave reflection does not play a dominant role. But if it does,
 the nice properties  of the phase-function will not
be guaranteed. Strong backscattering can have
three effects on the phase-function;
 (i) it may oscillate (often more rapidly than the solution to
 (\ref{waveq}) itself).  (ii) its amplitude may drop by several orders
 of magnitude (iii) it may have poles (see Appendix A of Andersson
 1993). It is clear that this may severely affect numerical
 calculations.

Consequently, the Andersson-Kokkotas-Schutz method must either be
adapted to this situation or used with some care. Here we have chosen
the second route. Thus the fact that we lose
numerical precision because of effects due to wave reflection in the exterior
limits our calculations somewhat. For example, for $M/R=0.44$ we are
restricted to ${\rm Im}\ \omega M < 2.5$ if we want six digit precision
in our final iterations.
This restriction is much less severe for less compact stars since the
surface of the star is then well outside the peak of the black-hole
curvature potential. In some cases we also find poles in the phase-function.
These can
masquerade as ``purely ingoing-wave modes'' for certain complex
frequencies. They are, of course, spurious and should not be taken
seriously.

However, these effects are not the main ones that limit the present
study. Due to numerical difficulties in  the interior calculation we
have been restricted to  ${\rm Im}\ \omega M < 1.25$ or so. The reason
for these difficulties is not well understood at the moment. This issue
must be studied in greater detail if more complete mode-surveys are to
be at all possible in the future. We have performed calculations using
both the equations of Detweiler and Lindblom (1985) and those of Ipser,
Price and Kojima (Ipser and Price 1991; Kojima 1992) for polar
perturbations. It seems that both sets suffer from these numerical
difficulties, although the latter formulation performs slightly
better.  In general, we have obtained identical numerical results from
the two formulations of the interior problem.

Finally, it is worth mentioning the possibility of anti-damped modes. A
linear instability of the star would correspond to an outgoing-wave
mode with a negative imaginary part of $\omega_n$. Our numerical
technique has allowed us to search for such modes and verify that they
do not exist. At least not in the part of the complex frequency plane
covered here; $-2< {\rm Im}\ \omega M < 0$.

\section{Results}

We present here the most extensive survey of gravitational-wave modes
for compact stars carried out to date. Our calculations cover stars
ranging in compactness from $R/M=100$, {\em i.e.,} something like a
hundred times as compact as a typical white dwarf, to ones very close
to the ultimate limit of stability in general relativity: $R/M = 9/4$.
The numerical results basically consist of a table describing how
various pulsation-mode frequencies change as the star is made more
compact. These results agree perfectly with our previous ones (Kojima,
Andersson and Kokkotas 1995) for the relevant cases.

In order to extract as much of the underlying physics as possible, it
is useful to consider these results in various units. For example, it
is natural to discuss the f-mode in terms of the density of the star
$\sim M/R^3$. It is well-known that the oscillation period of the
f-mode depends on the average density of the star. But at present we
have no such understanding of the gravitational-wave modes. One could,
for example, imagine that the periods of these modes will be associated
with the travel time for a gravitational wave that crosses the star,
and thus be related to the radius ($R$). Or maybe these modes are
similar to the quasinormal modes of black holes, and are best studied
in units of mass ($M$)? In fact, we will find all these units useful in
the following discussion.

\subsection{In units of density}

It seems natural to begin a discussion of the pulsation modes of
uniform density stars with the single mode that exists in Newtonian
theory. This is the f-mode and, as  mentioned above, it is best studied
in units of density. This mode has already been studied in some detail
by Detweiler (1975), and hence some of our results  may not be too
surprising. In Figure 1 we show how the complex frequency of the f-mode
changes as the star is made increasingly relativistic. For large values
of $R/M$ the oscillation frequency is very close to the Newtonian value
(\ref{fmode}). In fact, it differs from this value by less than 1\% for
stars less compact than $R/M \approx 5$.
 This means that the prediction of Newtonian theory is remarkably
 accurate also for neutron stars, even though these are very
 relativistic objects. As the star is made more compact the damping of
 the f-mode reaches a maximum. This feature was also noticed by
 Detweiler. We find that this maximum occurs for $R/M\approx 3.7$ and
 corresponds to ${\rm Im}\ \omega (R^3/M)^{1/2} \approx 4.6\times
 10^{-4}$. These values are in good agreement with Detweiler's results.
 After this maximum the damping of the f-mode decreases, as does the
 pulsation frequency.  Detweiler suggests that the complex frequency
becomes exactly zero at $R/M=9/4$. This would indicate that the f-mode becomes
secularly unstable in the extreme limit of relativity.


How are we to understand the maximum in the f-mode damping rate? A reasonable
answer is provided by Detweiler (1975). At first, as the star is made
more relativistic, emission of gravitational radiation becomes a more
efficient mechanism for energy release and thus the imaginary part of
$\omega_n$ increases. However, as $R/M$ decreases more and more of the
black-hole potential barrier is unveiled. Finally, at $R/M \approx 3$
the peak of the barrier emerges and the problem becomes analogous to
one of barrier scattering in quantum mechanics.  A gravitational wave
trying to escape to infinity will be partially reflected by the
barrier, and thus the damping time of the modes will be longer. This
effect is amplified by the fact that the oscillation frequency
decreases with $R/M$.

The idea that the potential barrier might be able to trap gravitational
waves led Chandrasekhar and Ferrari (1991b) to the discovery of axial
modes for very compact stars. That similar trapped polar modes exist
was recently shown by us (Kojima, Andersson and Kokkotas 1995). In
fact, we argued that the axial and polar modes behave in a  similar way
as the star becomes very compact. If so, it would seem likely that the
two sets of modes rely on the same physical mechanism for their
existence. In our view, both sets are gravitational-wave modes that
depend mainly on the character of the curved spacetime in the vicinity
of the star. In Figure 1 we  show the first  of these axial and
polar modes. As the star is made more compact the oscillation
frequencies and the damping rates for these modes generally decrease.

Again, that makes sense because of the increasing importance of the
potential barrier. In Figure 1 the axial and the polar mode seem to
behave quite differently for very compact stars, however. The damping
of the axial mode decreases monotonically, whereas the polar-mode
damping reaches a local minimum, a local maximum, and then falls off
rapidly in a way  similar to the f-mode damping rate.
The obvious question is, does this not contradict our previous
conclusions  (Kojima, Andersson and Kokkotas 1995) that were based on
the observed \underline{similarity} between the axial and the polar
modes?

The answer to this question is provided if one plots the real and the
imaginary parts of $\omega_n$ separately as functions of the
compactness of the star. An attempt to do this is Figure 2 of Kojima,
Andersson and Kokkotas (1995), but because of the method we used to
identify modes in that study the finer details could not be resolved.
When the data obtained by the recent calculations is used a remarkable
picture emerges. In Figure 2 it is clear that the polar pulsation mode
frequencies -- a set that includes the f-mode -- show a series of avoided
crossings (cf. for example Figure 17.7 in Cox 1980)
as the compactness increases.
Meanwhile, the axial spectrum does not have such features. This is a
totally unexpected behaviour. Even more surprising is the way in which
the ``agreement'' between axial and polar modes prevails. Notice, for
example, that the first axial w-mode is  very close to the first polar
w-mode for $M/R <0.439$.
 This was the kind of evidence on which we based our previous
 conclusions on the nature of these modes. But as the star is made more
 compact  a ``re-ordering'' occurs. The first axial mode becomes
 similar to the f-mode, while the first polar w-mode approaches the
 second axial mode. At first this seems very peculiar, but it does make
 sense if these modes are all considered as trapped inside the peak of a
 potential barrier. The two potentials governing axial and polar
 perturbations in the exterior vacuum are very similar (Chandrasekhar
 1983) and they should support similar sequences of trapped modes. The
 only surprising feature that remains is that {\em the f-mode must be
 considered on equal footing with the polar w-modes}.
 It is only the first in a sequence of trapped polar gravitational-wave
 modes. This is intriguing evidence for the richness of this problem.
 Equally interesting is the behaviour of the damping rates as the star
 gets more compact. At present we have no good explanation for the
 features in the damping rates that are apparent in Figure 2, however.


Finally, it may be worthwhile commenting on the fact that Detweiler
(1975) did not find the trapped w-modes. It seems likely that the
reason for this is that his method was based on a variational
principle. Thus it requires an initial guess for the f-mode
eigenfunction, and  modes that do not have similar eigenfunctions will
not be identified.

\subsection{In units of mass}

The present evidence supports the idea that the same physical
mechanism  gives rise to polar and axial gravitational-wave modes. In
fact, the fluid must  play a minor role since axial perturbations do
not couple to pulsations in the fluid here (Thorne and Campolattaro
1967). We must look to the curvature of spacetime if we want to explain
these modes. Then it is natural to wonder whether the w-modes are in
some sense similar to the quasinormal modes of black holes. These are
pure ``spacetime'' modes. In order to explore this issue, it is helpful
to display our results in units of mass.

Two things follow immediately from the present results for rapidly
damped modes: (i) {\em Polar and axial w-modes exist for all stellar
models} and (ii) {\em There are two distinct families of such modes.}
Both these results were anticipated from previous evidence (Kojima,
Andersson and Kokkotas 1995; Leins, Nollert and Soffel 1993).
Nevertheless, the present study brings these results beyond the realm
of mere possibilities.


The first result follows immediately from Figure 3, where we show the
extension of the data in Figure 1 for less relativistic stars. These
modes -- that make up the first family of w-modes --  approach $\omega
M = 0$ as $R/M \to \infty$. As the star becomes more compact $\vert
\omega_n M \vert$ reaches a maximum. This maximum is greater for the
higher overtones (larger $n$). After this maximum  $\vert  \omega_n M
\vert$ decreases until, for very compact stars, the behaviour is that
shown in Figure 1. The family of polar modes shown in Figure 3 is the
one discussed by Kokkotas and Schutz (1992).

The qualitative behaviour of the axial and polar modes in
Figure 3 is almost identical. If these were ``pure''
spacetime modes this similarity would be expected, especially since the
axial and the polar spectra are identical for black holes
(Chandrasekhar 1983). The present result therefore strongly supports
the idea that the stellar fluid is of very little importance for these modes.
Since these modes depend mainly on the spacetime curvature (this is
especially clear for extremely compact stars) we will from now on
refer to them as the ``curvature modes''.


We cannot see that the curvature modes in Figure 3 have anything
whatsoever in common with the modes of a black hole. The second family
of w-modes, shown in Figure 4, is somewhat more reminiscent of the
black-hole modes.  As $R/M \to 9/4$ these modes approach  constant complex
frequencies. These values are approximated in Table 1. It is
clear that these modes do not behave in the same way as the ones in
Figure 3 as we vary the compactness of the star. We can only find one
polar mode of the second type for all values of $R/M$
in the range $9/4 < R/M
\le 100$, but further ones (both axial and polar) exist for
sufficiently compact stars. These
new modes seem to emerge from the imaginary frequency axis.
We can find the first axial mode in the family for $M/R > 0.127$, and the
second polar one for $M/R > 0.227$. A second axial mode appears
for $M/R > 0.253$. The calculation becomes increasingly
difficult as ${\rm Im}\ \omega M$ increases and therefore we have only
been able to obtain partial results for the third polar mode in this
sequence. It seems to exist for $M/R > 0.299$. Anyway,
the results imply that only a (small) finite  number of such modes
exist for each star.  For reasons that will be discussed in more detail
in section 4.2 we will call this second set of modes the ``interface modes''.

As can be seen in Table 1, the interface modes do not have
a clear relation to the black-hole modes.  This is, of course, a
difficult conclusion to prove. We certainly do not expect the stellar
modes to approach the black hole ones as $R/M \to 9/4$.
 The star does not smoothly transform into a black hole in that limit.
 However, it is conceivable that these modes can be somehow associated
 with the peak of the exterior potential barrier for very compact stars
 (rather than the potential well, as the ``trapped'' modes of the first
 family). If so, these modes would be closely related to the black hole
 ones.


The behaviour of the first polar interface mode as $R/M \to
9/4$ is very peculiar. We do not profess to understand the ``wiggles''
that are apparent in Figure 3 at all.

\subsection{In units of radius}

Finally, we want to see whether the gravitational-wave modes discussed
here depend on the physical size of the star. This would, for example,
be the case if they depend on the time it takes a gravitational wave to
cross the star. This can easily be tested by displaying our results in
units scaled to the radius of the star. When we do that it is clear
that the pulsation frequencies of the curvature modes
discussed above do depend on the radius of the star. From Figure 5 it
follows that  ${\rm Re}\ \omega_n R$ approach constant values as
$R/M \to \infty$. Estimated values (for $R/M = 100$) are given in Table
2. In fact, the oscillation frequencies for more compact stars typically differ
from those in Table 2 by less that 1\% for $R/M > 60$. An interesting
observation is that the separation of the various pulsation frequencies
is not too different from  $\pi$. This fact is important for the
discussion of the physics of these modes in the following section.


\section{On the two families of w-modes}

In the previous section we presented extensive results for
gravitational-wave modes of uniform density stars. These results should
help us understand the origin and the nature of these modes.

The first obvious observation is that, as far as the gravitational-wave
modes are concerned, the axial and the polar spectra are very similar.
As we understand it, this implies that the spacetime curvature plays
the main role in this game. In fact, the qualitative similarity between
axial and polar modes provides an enormous advantage for this
discussion. The axial modes are described by a single wave-equation,
whereas the polar ones are governed by two coupled equations. Hence, it
is much easier to interpret the axial modes.

Any conclusions drawn from the present data must, of course, also agree
with previous results. We have found that  two  distinct
families of w-modes exist. This idea was first proposed by Leins,
Nollert and Soffel (1993), who found a few extra modes that had not
been identified by Kokkotas and Schutz (1992). That these modes do,
indeed, exist for polytropes has recently been verified by Andersson,
Kokkotas and Schutz (1995a). Leins {\em et al.} argue that the new modes
are distinct for two reasons. The first is the one that we have relied
upon in this investigation:
Modes belonging to different families behave differently as the
compactness of the star changes. The second argument is based on the
eigenfunctions inside the star. Leins {\em et al.} found that the
eigenfunctions of the w-modes of Kokkotas and Schutz were concentrated
at the centre of star whereas the new modes seemed localized at the
surface. We feel that an argument based on the eigenfunctions inside
the star must be used with some care, however. There is no apparent
reason why the eigenfunctions pertaining to the gravitational-wave
degrees of freedom could not be localized outside the star. In fact, it
seems likely that this is the case for the trapped modes that occur
when the star is very compact.

Neither of the above arguments for why the two families of modes are
distinct is really satisfactory, however. We must also understand the
physical differences between the two families. Based on the evidence
provided by the numerical results discussed here we suggest the
following: {\em There exist two different families of w-modes for both
axial and polar perturbations. One of these families is primarily
associated with the spacetime curvature inside the star.
The second family of modes is
mainly associated with the surface of the star} and arises because of
the ``discontinuity'' there. Below we offer arguments that support this
and indicate what we believe is the physics behind these modes. These
arguments need be supported by more detailed, preferably analytical,
work in the future.

\subsection{Modes associated with the spacetime curvature}

Let us first discuss the modes associated with the stellar interior.
These are the ones depicted in Figures 1-3 and 5 above. They were
discovered by Kokkotas and Schutz (1992). We suggest that these modes
arise, not because of the coupling between the spacetime and the fluid,
as has been suggested previously, but rather as gravitational waves
that are ``trapped'' by the spacetime curvature inside the star. It is
easy to see how this may happen if one plots the gravitational-wave
speed as measured by an observer at infinity: $e^\nu$ as a function of $r$.
This has a
minimum at the centre of the star, and the interior w-modes would be
trapped in this ``bowl'' of curvature. Moreover, such modes would
naturally be concentrated at the centre of the star, which agrees well
with the  eigenfunctions constructed by Leins, Nollert and Soffel
(1993). Hence, it makes sense to refer to these modes as ``curvature modes''.

It seems reasonable that the discontinuity provided by the
surface of the star will be able to partially reflect gravitational
waves, and therefore that the curvature modes will somehow be confined to
$r<R$. We intentionally use the word ``discontinuity'' in a very vague
sense here. In principle, some sort of discontinuity should always be
present at the stellar surface. In the case of uniform density stars
this is obvious, but also for more realistic stellar models, where the
density falls off towards the surface, will there be discontinuities (even
though these may appear in the higher derivatives of the relevant quantities).
More useful than an actual mathematical discontinuity is a rapid change
in the variables, for example the gravitational-wave speed, that can act as
an effective discontinuity for waves of certain wavelengths.

Anyway, a naive and useful argument leads to standing
wave solutions essentially of form $\sin(\omega r)$ inside the star.
If this were
the true form of the solutions, and the modes only leak out slowly
through the surface (so that we can assume a ``zero'' boundary
condition at the surface) we would get
\begin{displaymath}
\omega_n R =  n \pi \quad , \quad n = 1,2,...
\end{displaymath}
This argument is far too simplistic, but it is rewarding to find that
two of its predictions agree well with our results for the curvature modes:
 There would exist
an infinite sequence of such modes, and their pulsation frequencies
${\rm Re}\ \omega_n R$ would be separated by $\pi$. A similar
dependence on the size of the star can be infered for the modes in the
simple toy model of Kokkotas and Schutz (1986).

When the star is made increasingly compact, the curvature modes discussed here
clearly change character and should be considered as trapped in the
potential well that arises inside the black-hole potential barrier.  In
that regime the fluid f-mode should also be considered as a trapped
mode. The evidence for this from Figures 1 and 3 are clear.

\subsection{Modes associated with the surface of the star}

The existence of the second family of w-modes may be more directly
due to the discontinuity at the surface of the star. Then the
mode-eigenfunctions need not be localized in the star. Rather, these
modes would be similar to modes for acoustic waves scattered of a hard
sphere. One would typically expect such modes to be short-lived
compared to modes trapped inside the star. This agrees well with the
evidence of Figures 3 and 4. As Jensen (1989) has shown, the problem of
acoustic waves and a hard sphere can be approached analytically.
It is intriguing to find that the modes in that model problem are quite
similar our second family of w-modes. Only a finite number of modes
exist for each multipole $\ell$, and there may be purely imaginary
ones. The latter feature suggests that one should perhaps not be
surprised to find stellar modes ``emerge'' from points on the
imaginary-frequency axis as in Figure 4 here. In our opinion, this
evidence is compelling and makes the association of the second family
of w-modes and the interface at the stellar surface likely.
Hence, we have decided to call these modes ``interface modes''.
In principle, our conjecture that
only a finite number of these modes exist should be testable. But at
present numerical difficulties restrict calculations
to ${\rm Im}\ \omega M < 1.25$
or so. Much better numerical schemes, or other formulations of the
problem, are required
to test this prediction.

\section{Concluding remarks}

We have presented and discussed the results of an extensive survey of
gravitational-wave modes for the simplest possible stellar model; that
of uniform density. The results are truly exciting and contribute
considerably to our understanding of the gravitational-wave modes of
compact stars. Nevertheless, it is clear that  much work remains
in this area.

We have shown that the problem of axial perturbations has all the
essential features of the polar problem, as far as  gravitational waves
are concerned. This establishes the existence of axial modes for
\underline{all} stellar models,  contrary to what has long been the
accepted view (Thorne and Campolattaro 1967; Chandrasekhar and Ferrari
1991b). The present results show that there exists two distinct
families of gravitational-wave modes. We suggest that one of these
familiesi, the curvature modes, corresponds to waves trapped in the
interior of the star,
whereas the second, the interface modes,
is associated with the ``discontinuity'' at the
surface of the star. The first set of modes is similar to those of the
toy model proposed by Kokkotas and Schutz (1986). The latter set of
modes would be analogous to the acoustic modes associated with a hard
sphere (Jensen 1989). While there should exist an infinite number of
curvature modes, the number of interface modes may well be  finite and
possibly small.

Considering that the uniform density star can only support a single
fluid mode  -- the Kelvin f-mode -- the unveiled richness of the
spectrum  of gravitational-wave modes is remarkable. Indeed, it may be
anticipated that many  features of a more realistic star, such as
discontinuities associated with a crust and a superfluid stellar
interior, will also affect the w-modes and (probably) give rise to
further sets of such modes.

The present work raises  many questions that future work must
satisfactorily answer. For example,
although our suggestions for the physical nature of the two w-mode
families are plausible, they must be supported by more detailed
studies. Fortunately, the axial problem has all the
essential features of the w-modes, and is considerably simpler than the
polar problem. Furthermore, this problem can be approached semi-analytically,
by (say) the WKB method. Hence, the axial problem may prove invaluable
for future exploration of the nature of the w-modes.

A second issue that must be addressed by future work concerns the
excitation of stellar pulsation modes in various dynamical situations.
Such work is of extreme importance if we are to properly understand a
pulsating star as source of gravitational waves. Such waves may be
detectable within the near future. By detecting a gravitational wave
that carries the signature of the pulsation modes of a neutron star, we
can hope to probe not only the interior of the star but also the nature
of spacetime itself. This is undoubtedly an intriguing prospect.

\acknowledgments

We are grateful to Bernard Schutz for many useful discussions. NA thanks the
Aristotle University of Thessaloniki for generous hospitality. NA and KDK
also acknowledge an exchange program supported by the British Council and the
Greek GSRT.

\pagebreak

\begin{table}
\caption{Comparing the frequencies of interface modes
in the limit $R/M\to 9/4$ to the quasinormal modes of a Schwarzschild
black hole. For the black hole the axial and the polar spectra are
identical. The stellar modes are all for $M/R = 0.44$ but are not
expected to deviate much from these values in the true limit. All
entries are listed in units of $M^{-1}$. }
\begin{tabular}[t]{ccc}
polar & axial  & black hole \\
0.67+0.13i & 0.74+0.29i & 0.37+0.09i\\
0.79+0.57i & 0.87+0.73i & 0.35+0.27i\\
0.91+1.04i &            & 0.30+0.48i
\end{tabular}
\end{table}

\begin{table}
\caption{ The constant values approached by ${\rm Re}\ \omega_n R$  as
$R/M \to \infty$. These results  are for the curvature modes,
and specifically $R/M = 100$. But the results would typically differ with less
than 1\% as long as $R/M > 60$. It is interesting to note that the
separation between consecutive modes is quite close to $\pi$. }
\begin{tabular}[t]{ccccc}
$n$ & \multicolumn{2}{c}{polar modes} & \multicolumn{2}{c}{axial modes} \\
&${\rm Re}\ \omega_n R$ & ${\rm Re}\ (\omega_n-\omega_{n-1})R$ & ${\rm
Re}\ \omega_n R$ &  ${\rm Re}\ (\omega_n-\omega_{n-1})R$\\
1 & 3.64  &      & (1.50)  & \\
2 & 7.02  & 3.38 & 5.35  & (3.85) \\
3 & 10.27 & 3.25 & 8.65  & 3.30 \\
4 & 13.46 & 3.19 & 11.87 & 3.22
\end{tabular}
\end{table}

\pagebreak

\vspace*{5cm}

\begin{figure}[htb]
\epsfysize=7cm
\centerline{\epsffile {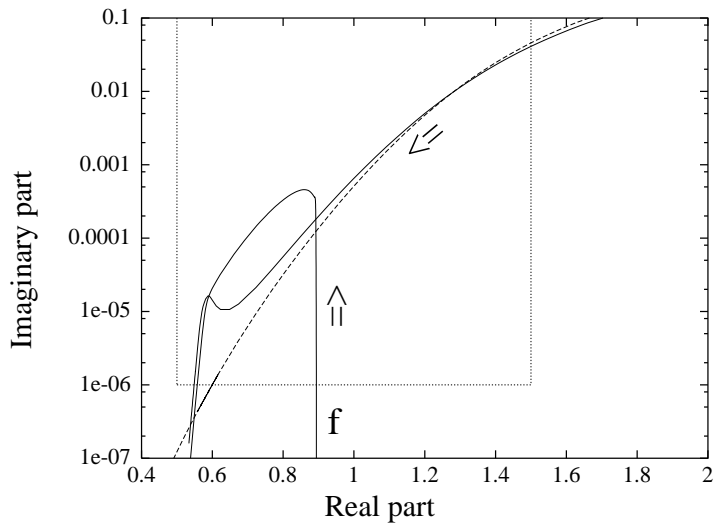}}
\caption{${\rm Im}\ \omega (R^3/M)^{1/2}$ vs ${\rm Re}\ \omega
(R^3/M)^{1/2}$ for the f-mode (solid and denoted by f). Also shown are
the first w-mode for polar (solid) and axial (dashed)
perturbations. Higher order modes behave in an almost identical way.
The box indicates the part of the frequency plane that
was covered in the study of Kojima, Andersson and Kokkotas (1995). That
study could not resolve the features in the lower left corner of that
box. Arrows indicate the direction of increasing compactness.}

\end{figure}

\pagebreak

\vspace*{5cm}

\begin{figure}[htb]
\epsfysize=7cm
\centerline{\epsffile {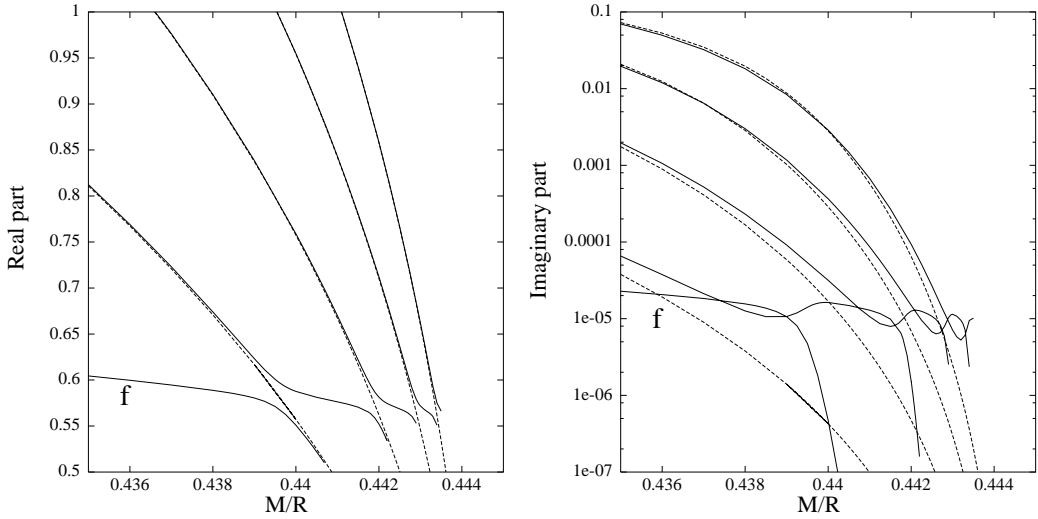}}
\caption{(left) ${\rm Re}\ \omega (R^3/M)^{1/2}$ and (right) ${\rm
Im}\ \omega (R^3/M)^{1/2}$ as functions of the compactness of the star
$M/R$ for the f-mode (solid and denoted by f), and the first few
w-modes for polar (solid) and axial (dashed) perturbations. Notice the
beautiful example of avoided crossings in the pulsation frequencies.}
\end{figure}

\pagebreak

\vspace*{5cm}

\begin{figure}[htb]
\epsfysize=7cm
\centerline{\epsffile {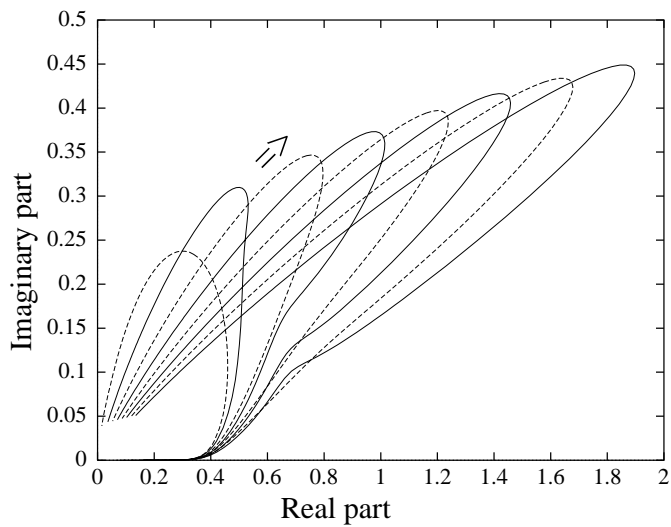}}
\caption{${\rm Im}\ \omega M$ vs ${\rm Re}\ \omega M$ for the
polar (solid) and axial (dashed) curvature modes. These modes are the
extension of Figure 1 for less compact stars. Although we only show the
first four axial and polar modes, it seems likely that an infinite
number of such modes exist. Note the remarkable qualitative agreement
between the two sets of modes. Arrows indicate the direction of
increasing compactness.}
\end{figure}

\pagebreak

\vspace*{5cm}

\begin{figure}[htb]
\epsfysize=7cm
\centerline{\epsffile {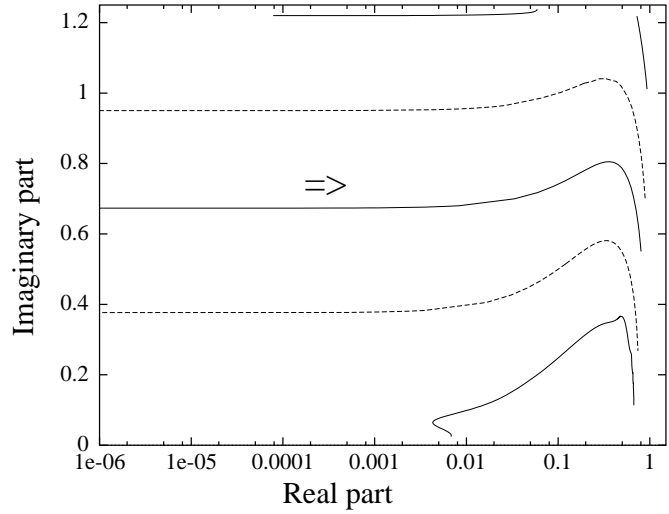}}
\caption{${\rm Im}\ \omega M$ vs ${\rm Re}\ \omega M$ for the
polar (solid) and  axial (dashed) interface modes. Only the first of these
modes exist for all values of  stellar compactness. Further modes arise
for sufficiently compact stars. The numerical calculation becomes
difficult for large ${\rm Im}\ \omega M $ and consequently we have only
partial data for the third of the polar modes modes. Arrows indicate the
direction of
increasing compactness.}
\end{figure}

\pagebreak

\begin{figure}[htb]
\epsfysize=7cm
\centerline{\epsffile {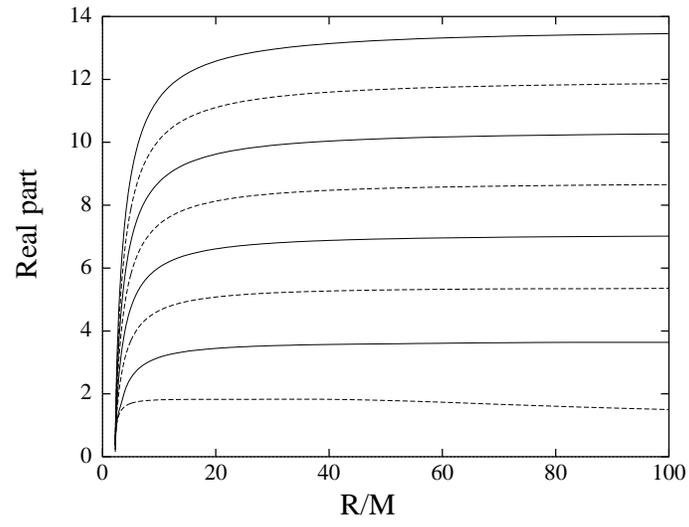}}
\caption{ ${\rm Re}\ \omega R$ vs $R/M$ for the polar (solid) and
axial (dashed) curvature modes.}
\end{figure}

\end{document}